\documentclass[aps,prl,twocolumn,groupedaddress]{revtex4-1}
\usepackage{amssymb,amsmath}
\usepackage{graphicx}
\usepackage{hyperref}

\begin{document}

\title{U(1) Emergence versus Chiral Symmetry Restoration in the Ashkin Teller Model}

\author{Soumyadeep Bhattacharya}
\email{sbhtta@imsc.res.in}
\affiliation{Institute of Mathematical Sciences, CIT Campus, Taramani, Chennai 600113, India}

\date{\today}

\begin{abstract}
We show that suppression of vortices in the Ashkin Teller ferromagnet on a
square lattice splits the order-disorder transition and opens up an
intermediate phase where the macroscopic symmetry enhances to U(1).
When we selectively suppress the formation of non-chiral vortices, chiral
vortices proliferate and replace the U(1) phase with a new phase where chiral
symmetry is restored.
This result demonstrates a fascinating phenomenon in which the symmetry
information encoded in topological defects manifests itself in the
symmetry of the phase where the defects proliferate.
We also show that this phenomenon can occur in all $\mathbb{Z}_n$ ferromagnets
with even values of $n$.
\end{abstract}


\maketitle
Topological defects play a crucial role in enhancing the microscopic
$\mathbb{Z}_n$ symmetry of discrete ferromagnets to a U(1) symmetry at the
macroscopic scale.
Proliferation of domain wall defects results in the formation of numerous
domains and the spins are able to change their orientation by arbitrary amounts
over large distances in a manner such that the macroscopic order parameter
exhibits angular fluctuations uniformly along all directions~\cite{einhorn1980physical}.
In some ferromagnets, however, discrete vortices proliferate simultaneously with
domain walls and disorder the system before the U(1) symmetry emerges.
A forced suppression of vortices, in such cases, can delay their proliferation
and allow the intermediate U(1) phase to manifest itself.

On the square lattice, $\mathbb{Z}_n$ ferromagnets with $n \geq 5$ exhibit an intermediate
U(1) phase without vortex
suppression~\cite{jose1977renormalization,elitzur1979phase,einhorn1980physical,
frohlich1982massless,lapilli2006universality,baek2009true,van2011discrete,
ortiz2012dualities,borisenko2012phase}.
The direct order-disorder phase transition in the $\mathbb{Z}_3$ (three state
Potts) ferromagnet, on the other hand, was recently shown to split under strong
suppression of vortices and an intermediate U(1) phase was uncovered~\cite{bhattacharya2015quasi}.
The $\mathbb{Z}_4$ Ashkin Teller ferromagnet also undergoes a direct order-disorder
transition~\cite{baxter2007exactly}.
The interplay between vortices and domain walls at this
transition is known to generate a line of continuously varying critical exponents~\cite{kadanoff1979multicritical}.
A decoupling of this interplay and demonstration of an extended phase where a microscopic $\mathbb{Z}_4$ symmetry
enhances to U(1) has interesting implications for quantum phase transitions in
antiferromagnets~\cite{quantum1,quantum2,quantum3}, melting of crystal
films~\cite{einhorn1980physical} and adsorption of gas particles on
metal surfaces~\cite{binder1982theoretical,hardsquare1,hardsquare2,hardsquare3,hardsquare4}.
Can the suppression of vortices open up a U(1) phase in this model as well?

In this article, we show that suppression of vortices in the $\mathbb{Z}_4$
ferromagnet on a square lattice indeed destroys the direct order-disorder
transition and results in the formation of an intermediate phase with emergent
U(1) symmetry.
Interestingly, however, U(1) emergence is not the only possibility for the
intermediate phase.
Due the even parity of $n$ in this model, we are able to distinguish between
chiral vortices and non-chiral vortices.
We show that a selective suppression of non-chiral vortices leaves the chiral
vortices to proliferate in the intermediate phase.
As a result, enhancement of the $\mathbb{Z}_4$ symmetry to a U(1) symmetry is
replaced by restoration of chiral symmetry, i.e. a $\mathbb{Z}_2$ subgroup
of the $\mathbb{Z}_4 \rightarrow \mathbb{Z}_2 \times \mathbb{Z}_2$ symmetry
is restored.
In order to verify that this phenomenon is not an artifact of the
$\mathbb{Z}_4$ symmetry, we demonstrate U(1) emergence versus chiral symmetry
restoration via vortex suppression in the $\mathbb{Z}_6$ ferromagnet as well.
We conclude that this phenomenon can occur in all ferromagnets with even $n$.

\begin{figure}
\includegraphics[width=0.8\linewidth]{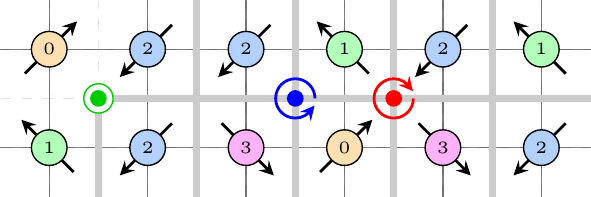}
\caption{\label{fig1}
(Color online)
A $\mathbb{Z}_4$ spin configuration is shown with domain walls (gray),
a non-chiral vortex (blue), a non-chiral antivortex (red) and a
chiral vortex (green).}
\end{figure}

In a general $\mathbb{Z}_n$ ferromagnet, a spin $s_i$ is placed at each vertex
$i$ of a lattice $\Lambda$ (a square lattice in this case) and each spin can be
in one of $n$ different states $s_i \in \{0,1,\ldots,n-1\}$ (Fig.~\ref{fig1}).
In the ordered phase a majority of the spins take up a common state while the
states are taken up arbitrarily in the disordered phase.
The vector order parameter which captures a direct transition between these two
phases is defined for a system of $L^2$ spins as $m \equiv (m_x,m_y)$ where
$m_x = L^{-2}\sum_i \cos 2 \pi s_i / n$ and $m_y = L^{-2}\sum_i \sin 2 \pi s_i / n$
~\cite{baek2009true,borisenko2012phase}.
The macroscopic symmetry of the system is reflected in the distribution
$P(m_x,m_y)$ of this order parameter.

Domain walls and vortex defects reside on the dual lattice $\Lambda'$, which in
this case is another square lattice shifted from $\Lambda$ by half a lattice spacing
along each direction~\cite{einhorn1980physical,ortiz2012dualities}.
If two neighboring spins on $\Lambda$ are in dissimilar states, then a domain wall
is placed on the edge of $\Lambda'$ separating the two spins (Fig.~\ref{fig1}).
Each vertex $i'$ in $\Lambda'$ is assigned a winding number
\begin{eqnarray}
\omega_{i'} = (\Delta_{b,a}^n + \Delta_{c,b}^n + \Delta_{d,c}^n + \Delta_{a,d}^n)/ n
\end{eqnarray}
which is essentially the finite difference equivalent of a circuit integral in
continuum space when each elementary square plaquette in $\Lambda$ is chosen as
a circuit and the spins at the four corners of each plaquette are in states
$a$, $b$, $c$ and $d$ when traversed in an anticlockwise sequence.
A vortex is present at $i'$ if $\omega_{i'} = +1$ and an anti-vortex is present if
$\omega_{i'} = -1$ (Fig.~\ref{fig1}).

$\Delta_{a,b}$ represents the difference, modulo $n$, between states $a$ and $b$.
More precisely, this modular difference is calculated as
\begin{eqnarray}
  \Delta_{a,b}^n =
  \begin{cases}
    a-b-n,& \text{if } a-b >    +n/2\\
    a-b+n,& \text{if } a-b \leq -n/2\\
    a-b,  & \text{otherwise}
  \end{cases}
\label{eqn_state_difference}
\end{eqnarray}
The asymmetry in this calculation is evident from the presence of an equality
condition in $a - b \leq -n/2$.
If a spin state $s_i$ is viewed as two dimensional unit vector oriented at an
angle $\theta_i = 2 \pi s_i / n$, then this calculation restricts the angle
difference to lie in $(-\pi,+\pi]$~\cite{bittner2005vortex,ortiz2012dualities}.
The asymmetry induced by the inclusion of $+\pi$ and exclusion of $-\pi$ is the
origin of chirality in vortices.
Consider, for example, a configuration of $\mathbb{Z}_4$ spins with states 0,1,2
and 2 arranged at the corners of a square plaquette when read in an anticlockwise
sense (Fig.~\ref{fig1}).
The winding number for this plaquette, according to the above calculation is +1,
indicating the presence of a vortex.
If, on the other hand, we calculate the winding number by traversing the plaquette
in a clockwise sense, then the winding number turns out to be zero, indicating the
absence of vorticity.
As we will show, this chirality that is built into the definition of vorticity can
have fascinating consequences for the phase diagram of the system.

In order to highlight these consequences, we define a different type of vorticity
which does not exhibit this chirality.
A non-chiral version of the modular difference is formulated as
\begin{eqnarray}
  \Delta_{a,b}^{n,nc} =
  \begin{cases}
    a-b-n,& \text{if } a-b > +n/2\\
    a-b+n,& \text{if } a-b < -n/2\\
    a-b,  & \text{otherwise}
  \end{cases}
\label{eqn_nc_state_difference}
\end{eqnarray}
The corresponding calculation for the winding number is
\begin{eqnarray}
\omega_{i'}^{nc} = (\Delta_{b,a}^{n,nc} + \Delta_{c,b}^{n,nc} + \Delta_{d,c}^{n,nc} + \Delta_{a,d}^{n,nc})/ n
\end{eqnarray}
The configuration of $\mathbb{Z}_4$ states 0,1,2 and 2, for example,
results in zero vorticity with this definition when calculated either in clockwise
or in anticlockwise sense.
Vortex defects with $\omega_{i'}^{nc} \neq 0$ are termed as non-chiral
vortices.
The special defects which have $\omega_{i'}^{nc} = 0$ but $\omega_{i'} \neq 0$ are
termed as chiral vortices.

The formation of vortices and antivortices can be suppressed by raising their core
energy by an amount $\lambda$~\cite{bhattacharya2015quasi}.
With the inclusion of such a term for suppressing
standard vortices, the clock Hamiltonian for a $\mathbb{Z}_n$ ferromagnet becomes
\begin{eqnarray}
\mathcal{H} = -\sum_{\langle i,j \rangle \in \Lambda} \cos 2\pi\left(s_i - s_j\right)/n
              + \lambda \sum_{i' \in \Lambda'} |\omega_{i'}|
\label{eqn_zn_clock_hamiltonian}
\end{eqnarray}
Non-chiral vortices can be selectively suppressed by replacing the second term
with $\lambda^{nc} \sum_{i' \in \Lambda'} |\omega_{i'}^{nc}|$.

We have simulated both these cases across a wide range of suppression strength
and temperature $T$ for different system sizes.
In our simulation, spins were initialized to a completely ordered configuration
and updated using the Metropolis single spin-flip algorithm as the plaquette based
term for vortex suppression cannot be incorporated into cluster update
algorithms~\cite{landau2014guide}.
The autocorrelation time for observables, which tend to be quite large for single
spin-flip algorithms, was measured at each parameter point.
Around $10^4$ uncorrelated configurations were discarded for equilibriation
following which measurements were made on $10^5$ uncorrelated configurations.

For each configuration, we calculated the magnetization
$|m| = \sqrt{m_x^2 + m_y^2}$, the angle of the order parameter $\phi = \arctan(m_y/m_x)$,
the density of domain walls $\rho_{dw}$ defined as the fraction of edges in
$\Lambda'$ containing a domain wall and the density for the different types of
vortex defects, defined as the fraction of vertices in $\Lambda'$ containing
a defect of the given type.
$\rho_{vx}$ represents the density of standard vortices with $\omega_{i'} \neq 0$,
$\rho_{ncvx}$ represents the density of non-chiral vortices and $\rho_{cvx}$ represents
the density of chiral vortices.

\begin{figure*}
\includegraphics[width=\linewidth]{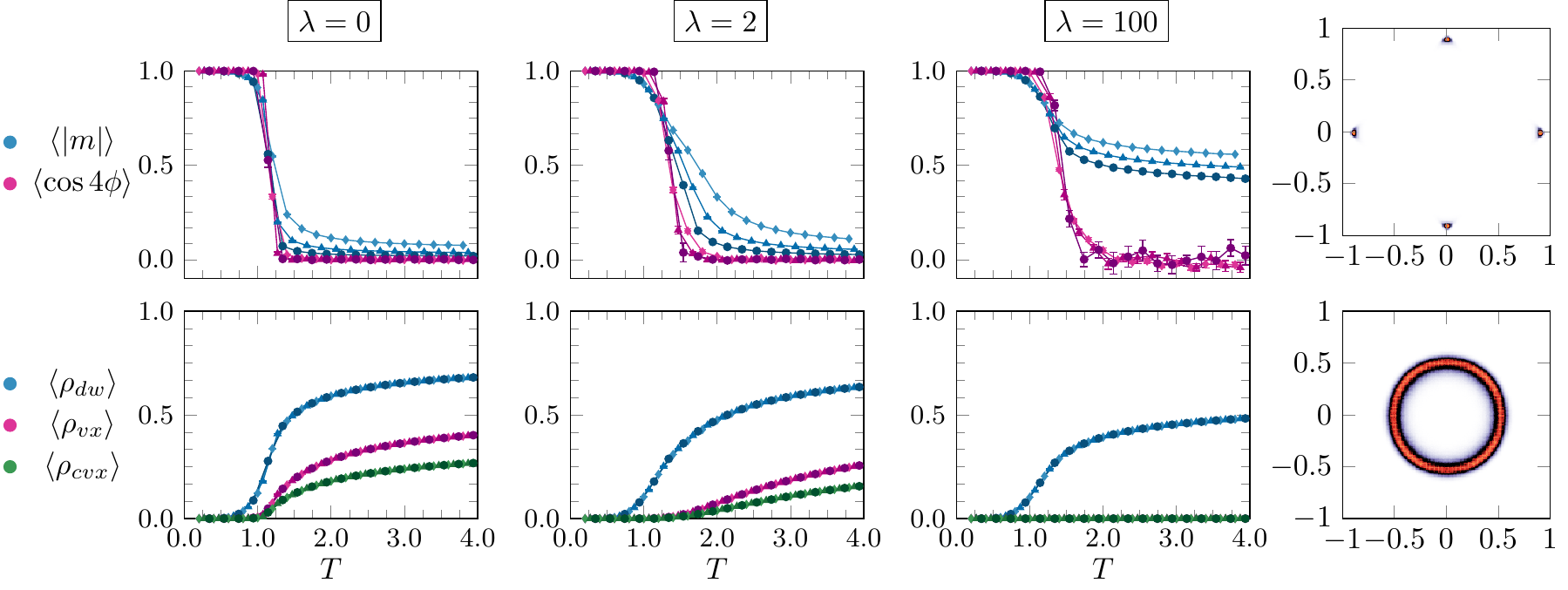}
\caption{\label{fig2}
(Color online) 
A split in the order-disorder transition and the appearance of an intermediate
phase is shown to occur with increasing suppression strength $\lambda$. 
Observable values correspond to system sizes $L=16$ (diamond), $L=32$ (triangle)
and $L=64$ (circle). For $\lambda = 100$, vortices are nearly absent and only
the domain walls are left to proliferate. The order parameter distribution
$P(m_x,m_y)$ for a $32^2$ system at this $\lambda$ shows $\mathbb{Z}_4$ symmetry
breaking in the ordered phase at $T=1.1$ (top right) and enhancement to U(1)
symmetry at $T=3.0$ (bottom right).}
\end{figure*}

\begin{figure}
\includegraphics[width=\linewidth]{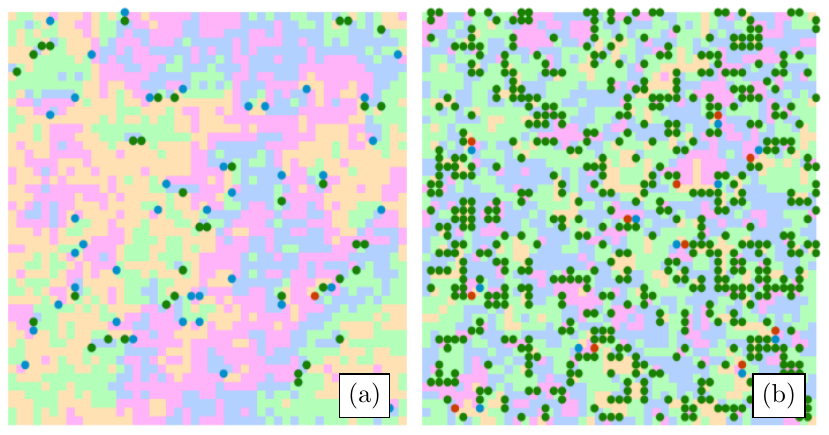}
\caption{\label{fig3}
(Color online)
Typical spin configurations for a $48^2$ system 
show (a) numerous domains and a few vortices in the emergent U(1) phase
when standard vortices are suppressed, and (b) proliferation of
chiral vortices in the chiral symmetry restored phase when non-chiral
vortices are suppressed.
Chiral vortices (green dots) are overlaid on the standard vortices
(red and blue) in (a) and shown alongside non-chiral vortices in
(b). Both configurations were generated at $T=2.5$ with a suppression value of 4.}
\end{figure}

In the pure $\mathbb{Z}_4$ clock ferromagnet ($\lambda = 0$), the magnetization
decays to zero across the direct order-disorder phase transition at $T \sim 1.1$ (Fig.~\ref{fig2}).
The pattern of symmetry breaking in the low temperature phase is captured by
$\langle \cos 4 \phi \rangle$ which takes a value +1 when the $\mathbb{Z}_4$
symmetry is broken.
Both domain walls and vortices are observed to proliferate simultaneously near
this transition, as indicated by an increase in their densities.
When the formation of vortices is suppressed ($\lambda = 2$), the transition
splits into two as indicated by the appearance of a two step decay in the
magnetization.
The first decay is accompanied by the proliferation of domain walls while the
second one corresponds to the disordering transition driven by the proliferation of
vortices.
This behavior is similar to the one obtained with vortex suppression in the $\mathbb{Z}_3$
ferromagnet~\cite{bhattacharya2015quasi}.
Spin configurations in the intermediate phase (Fig.~\ref{fig3}(a)) show a few bound pairs of
vortices and fragmented domains typical of quasi long range order~\cite{bhattacharya2015quasi}.
This intermediate phase can be extended further by pushing the disordering transition
to higher temperatures via strong suppression of vortices ($\lambda = 100$).
When the formation of vortices is completely forbidden, the disordering transition
vanishes and the system is left with a single transition from the $\mathbb{Z}_4$
symmetry broken phase to the emergent U(1) phase(Fig.~\ref{fig2}).
This result is interesting because a thermodynamically stable emergence of U(1) symmetry,
which is usually reported only at the disordering transition point, is shown to appear
throughout an extended phase in this case.
However, as we show next, U(1) emergence is not the only possibility for the
intermediate phase.

\begin{figure*}
\includegraphics[width=\linewidth]{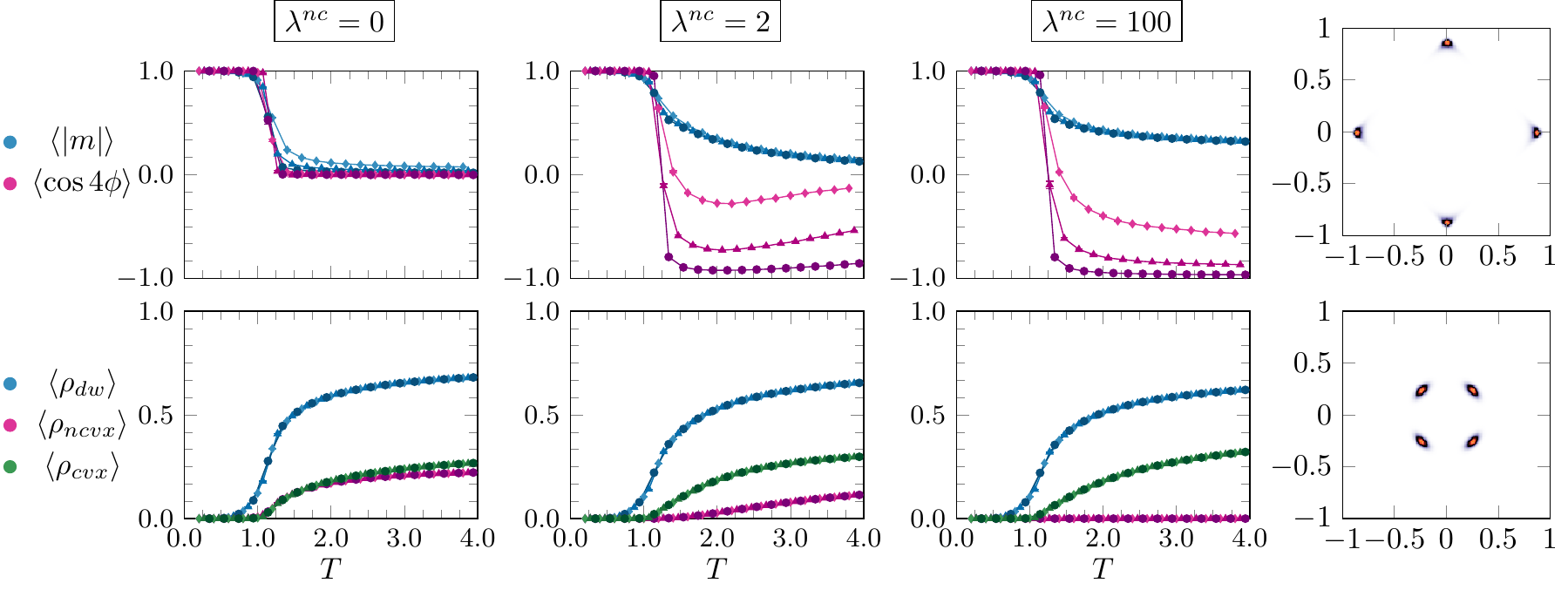}
\caption{\label{fig4}
(Color online)
The order-disorder transition is replaced by a chiral symmetry restoring
transition when non-chiral vortices are suppressed using increasing
values of $\lambda^{nc}$.
The $\mathbb{Z}_4$ symmetry breaking at $T = 1.1$ in the ordered phase (top right)
and chiral symmetry restoration at $T = 3.0$ (bottom right) are clearly
reflected in the order parameter distribution obtained for a $32^2$ system
with $\lambda^{nc} = 100$.
System sizes used to calculate observable statistics are the same as those
in Fig.~\ref{fig2}.}
\end{figure*}

\begin{figure}
\includegraphics[width=\linewidth]{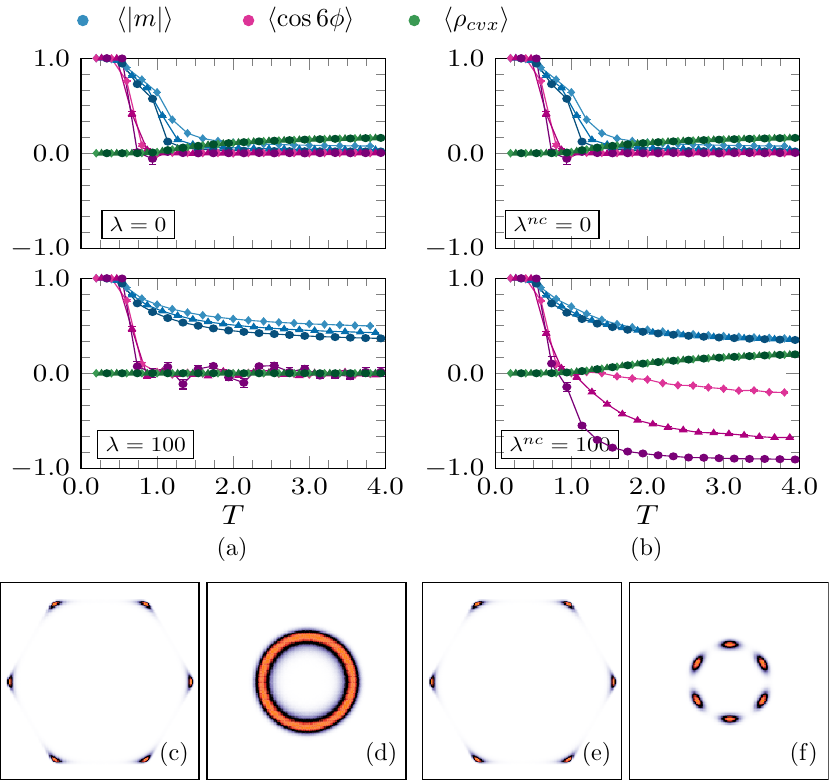}
\caption{\label{fig5}
(Color online)
The emergent U(1) phase in the $\mathbb{Z}_6$ clock ferromagnet
is (a) shown to extend upon suppression of standard vortices, and
(b) shown to get replaced by a chiral symmetry restored phase upon
suppression of non-chiral vortices.
The order parameter distribution shows that the $\mathbb{Z}_6$
symmetry is broken in the ordered phase in both cases of suppression
(c) and (e), but gets enhanced to U(1) in the former (d), and replaced
by chiral symmetry restoration in the later (f).}
\end{figure}

We return to the pure $\mathbb{Z}_4$ ferromagnet and begin a selective suppression
of non-chiral vortices using a $\lambda^{nc}$ term for (\ref{eqn_zn_clock_hamiltonian}).
When $\lambda^{nc} = 0$, both chiral and non-chiral vortices proliferate
simultaneously accompanied by the proliferation of domain walls across the
direct order-disorder transition (Fig.~\ref{fig4}).
When the non-chiral vortices are suppressed ($\lambda^{nc} = 2$), the transition
appears to split again.
In this case, however, the proliferation of domain walls is closely followed by the
proliferation of chiral vortices.
With increasing $\lambda^{nc}$, the proliferation of both domain walls and chiral
vortices remains unchanged while proliferation of non-chiral vortices is observed
to weaken.
The crucial difference in the pattern of symmetry breaking on the high temperature
side is captured by $\langle \cos 4 \phi \rangle$, which changes its sign to a negative
value.
This represents the restoration of $\mathbb{Z}_2$ chiral symmetry and indicates a
transition from order to incomplete order~\cite{todoroki}.
Examples of similar $\mathbb{Z}_2$ symmetry restored phases are the $\langle
\sigma \rangle$ phase of the Ashkin Teller model obtained beyond the clock point
in three dimensions~\cite{ditzian}
and the disordered flat phase in restricted solid-on-solid models of crystal
growth~\cite{zippering}.
Typical spin configurations obtained for this phase clearly reveal the
proliferation of chiral vortices (Fig.~\ref{fig3}(b)).
Interestingly, the full $\mathbb{Z}_n$ symmetry is never restored at any value of
temperature above the chiral symmetry restoring transition.
Instead, $\langle \cos 4 \phi \rangle$ appears to become more negative with
increasing $L$ (Fig.~\ref{fig4}), indicating the persistence of a thermodynamically stable chiral
symmetry restored phase even at high temperatures.
This result suggests that proliferation of non-chiral vortices is not strong
enough to drive a disordering transition and accompanies a gradual crossover
to disorder instead.

The restoration of a $\mathbb{Z}_2$ symmetry raises an important question: is
this phenomenon a special feature restricted to the $\mathbb{Z}_4$ ferromagnet,
possibly because it can be decomposed into a $\mathbb{Z}_2 \times \mathbb{Z}_2$
model?
While defining the non-chiral vortices, we had not restricted ourselves to $n = 4$.
The distinction between chiral and non-chiral vortices was made possible
using the asymmetry of modular differences for the particular case when
$\Delta_{a,b} = -n/2$ in (\ref{eqn_nc_state_difference}).
Such cases appear when $n$ is even.
It should, therefore, be possible to verify whether the replacement of U(1)
emergence by chiral symmetry restoration occurs in models with higher even
values of $n$ as well.

We have simulated the $\mathbb{Z}_6$ clock ferromagnet on the square lattice
with different types of vortex suppression.
This model exhibits a narrow intermediate U(1) phase even without the
suppression of vortices~\cite{lapilli2006universality,ortiz2012dualities,borisenko2012phase}.
When we suppress the formation of the standard vortices, the U(1) phase is
observed to extend further upto higher temperatures and chiral vortices
are nearly absent in the phase (Fig.~\ref{fig5}).
On suppression of non-chiral vortices, however, the chiral vortices
are left behind to proliferate and they drive a phase transition from the
ordered phase to a chiral symmetry restored phase.
The symmetry restoration is clearly reflected in the order parameter
distribution and in $\langle \cos 6 \phi \rangle$ which changes sign
across the transition.
This result confirms that chiral symmetry restoration is realizable in general
$\mathbb{Z}_n$ ferromagnets with even values of $n$.

The subtle difference between the proliferation of standard vortices, which
contains both chiral and non-chiral defects, and the proliferation of
chiral defects highlights the crucial role played by these topological defects
in determining the phase diagram of $\mathbb{Z}_n$ ferromagnets.
Our result can have interesting implications for phase transitions in a variety
of other systems like superfluids and superconductors which are also governed
by the proliferation of vortex defects~\cite{kohring1986role,shenoy1990enhancement}.
Moreover, our demonstration provides
a simple example of how the symmetry information contained inside topological
defects can determine the symmetry of the phase in which they proliferate.
It would, therefore, be interesting to apply our technique and open up richer
phase diagrams for systems like liquid crystals~\cite{lammert1993topology} in
which the defects are associated with higher symmetries.


\end{document}